\documentclass[twocolumn,showpacs,showkeys]{revtex4}
\usepackage{graphicx}
\usepackage{bm}
\usepackage{color}
\usepackage{amsmath}
\usepackage{natbib}

\begin{document}

\title{Hydrodynamics of quantum corrections to the Coulomb interaction via the third rank tensor evolution equation:
Application to the Langmuir waves and the spin-electron-acoustic waves}

\author{Pavel A. Andreev}
\email{andreevpa@physics.msu.ru}
\affiliation{Faculty of physics, Lomonosov Moscow State University, Moscow, Russian Federation, 119991.}

\date{\today}

\begin{abstract}
If we study the quantum effects in plasmas in terms of traditional hydrodynamics via the continuity and Euler equations
we find the quantum Bohm potential and the force of spin-spin interaction.
However, if we extend the set hydrodynamic equations beyond the 13-moments approximation,
and include the third rank tensor evolution equation along with the pressure evolution equation,
we obtain the quantum corrections to the Coulomb interaction.
It is found in contrast with the fact that hydrodynamic equations for the higher rank tensors do not contain interaction
in the classic plasmas studied in the selfconsistent (meanfield) approximation.
Therefore, we present the quantum hydrodynamic model,
where the quantum effects are studied beyond the quantum Bohm potential.
Developed model is considered in two regimes:
all electrons in plasmas are considered as the single fluid,
and the separate spin evolution regime,
where electrons with different spin projections are considered as two different fluids.
To illustrate the fundamental meaning of found quantum effects
we demonstrate their contribution in the spectrum of the Langmuir waves and the spin-electron-acoustic waves.
It is worth to mention that the application of the pressure evolution equation
ensures that the contribution of pressure in the Langmuir wave spectrum is proportional to $(3/5)v_{Fe}^{2}$,
unlike $(1/3)v_{Fe}^{2}$ appearing from hydrodynamics based on the continuity and Euler equations,
where $v_{Fe}$ is the Fermi velocity.
Same correction corresponds on other plasmas phenomena like the speed of sound for spin-electron-acoustic waves.
Moreover, it is found that novel quantum effects provide the novel wave solutions.
\end{abstract}

\pacs{}
\keywords{quantum hydrodynamics, pressure evolution, separate spin evolution, waves}


\maketitle


\section{Introduction}

Passed two decades the quantum plasmas is intensively studied \cite{Kremp PRE 99}, \cite{Shokri PoP 99}, \cite{Golubnychiy PRE 01}.
Quantum effects in electron component of plasmas becomes noticeable at low temperature,
hence, mostly, electrons are in the state of degenerate electron gas.
In 1999-2006 community was mostly focused on the quantum Bohm potential
\cite{Haas PRE 00}, \cite{Anderson PRE 02}, \cite{Haas PoP 03}, \cite{Haas PoP 05}, \cite{Maksimov QHM 99},
which can be interpreted as the quantum part of pressure appearing in addition to the Fermi pressure.
The quantum Bohm potential increases the contribution of the Fermi pressure.
This increase becomes comparable with the Fermi pressure
if the wavelength decreases down to average interparticle distance.
Some of these effects are described in the following reviews \cite{Shukla UFN 10}, \cite{Shukla RMP 11}.

Spin dynamics of electrons gives diverse effects in quantum plasma behavior.
Fundamental derivation of quantum hydrodynamic equations for spin quantum plasmas is made in 2001
\cite{MaksimovTMP 2001}, \cite{MaksimovTMP 2001 b}.
However, major interest to this field appeared after 2007
\cite{Andreev VestnMSU 2007}, \cite{Marklund PRL07}, \cite{Brodin NJP 07}, \cite{Andreev AtPhys 08}, \cite{Mahajan PRL 11}, \cite{Koide PRC 13},
(see also \cite{Uzdensky RPP 14}).

Presence of the spin of electron leads to additional hydrodynamic equation.
It is the spin density evolution equations.
Quasi-classic part of the flux of spin is considered in all papers on the spin quantum plasmas.
Quantum part of the flux of spin,
which is the analog of the quantum Bohm potential is considered in some papers (see for instance \cite{Mahajan PRL 11}).
However, the pressure-like part of the flux of spin, called the Fermi spin current or the thermal spin current,
is not considered in majority of works in this field.
Its explicit form is considered in recent papers for degenerate electrons
\cite{Andreev PTEP 19}, \cite{Andreev PoP kinetics 17 a}, \cite{Andreev PoP kinetics 17 b}, \cite{Andreev PoP 16 sep kin}.

The separate spin evolution quantum hydrodynamics,
where electrons are considered as two different fluids is developed in 2015 \cite{Andreev PRE 15 SEAW}, \cite{Andreev AoP 15 SEAW}.
This model shows that there is the spin-electron-acoustic wave in the spin polarized electron gas
\cite{Andreev PRE 15 SEAW}, \cite{Andreev EPL 16}, \cite{Andreev APL 16}.
Its existence is caused by the difference of pressures for the spin-up and spin-down electrons.

There are examples of extended hydrodynamics \cite{Tokatly PRB 99}, \cite{Tokatly PRB 00}, \cite{Miller PoP 16},
where the equations for the evolution of the second rank tensors (the momentum flux and the spin flux) are included.
It covers some spin related wave phenomena in quantum plasmas.

Here, I demonstrate that the account of the higher rank material field tensors like the momentum flux and the third order tensor,
which is the flux of the momentum flux, leads to new source for quantum effects in the quantum hydrodynamics of plasmas.
It is true for the spin-less regime
since it appears as the quantum part of the Coulomb interaction.
It also gives contribution in the spin-electron acoustic waves.

Obviously, similar generalization can be made for the spin-spin interaction.
Moreover, the third rank tensor, which is the flux of the spin-current,
evolution equation can be also considered for complete model of spin effects.
These generalizations are left for future papers.

This paper is organized as follows.
In Sec. II some fundamental definitions are introduced.
Final equations for the suggested model are also presented in Sec. II.
In Sec III dispersion dependencies are derived and analyzed.
In Sec. IV a brief summary of obtained results is presented.

\section{Electrostatic limit of the extended separate spin evolution hydrodynamics}

Here, I present the 20-moments quantum hydrodynamics,
which demonstrates novel source of quantum effects generalizing the quantum Bohm potential contribution.

First equation of all sets of hydrodynamic equations is
the continuity equation demonstrating the conservation of the number of particles:
\begin{equation}\label{QPhrt20 cont eq via v} \partial_{t}n_{s}+\nabla\cdot (n_{s}\textbf{v}_{s})=0. \end{equation}

The velocity field $\textbf{v}$ presented in the continuity equation obeys the Euler equation,
which has the following form for bosons in the BEC state
$$mn_{s}\partial_{t}v_{s}^{\alpha} +mn_{s}(\textbf{v}_{s}\cdot\nabla)v_{s}^{\alpha}
+\partial_{\beta}T_{s}^{\alpha\beta}$$
\begin{equation}\label{QPhrt20 Euler}
+\partial_{\beta}p_{s}^{\alpha\beta}=-q_{e}n_{s}\partial^{\alpha} \Phi.\end{equation}

The electrostatic potential $\Phi$ on the right-hand side of equation (\ref{QPhrt20 Euler})
has the following explicit form:
\begin{equation} \label{QPhrt20 int pot def}  \Phi(\textbf{r},t)=q_{e}\int d\textbf{r}'
\frac{1}{|\textbf{r}-\textbf{r}'|}(n_{\uparrow}(\textbf{r}',t)+n_{\downarrow}(\textbf{r}',t)-n_{0i}). \end{equation}

Quasi-electrostatic potential $\Phi(\textbf{r},t)$ (\ref{QPhrt20 int pot def}) obeys the Poisson equation
\begin{equation} \label{QPhrt20 int pot PE} \triangle\Phi=-4\pi q_{e}(n_{\uparrow}+n_{\downarrow}-n_{0i}),\end{equation}
where $n_{0i}$ is the equilibrium concentration of ions.

The left-hand side of the Euler equation contains the tensor associated with the quantum Bohm potential $T_{s}^{\alpha\beta}$.
The noninteracting part of the quantum Bohm potential is given by equation
\begin{equation} \label{QPhrt20 Bohm tensor single part}
T_{s}^{\alpha\beta}=-\frac{\hbar^{2}}{4m}\biggl[\partial_{\alpha}\partial_{\beta}n_{s}
-\frac{\partial_{\alpha}n_{s}\cdot\partial_{\beta}n_{s}}{n_{s}}\biggr].\end{equation}
Deviations from the pressure and the noninteracting part of the quantum Bohm potential appear together $p_{s}^{\alpha\beta}+T_{s}^{\alpha\beta}$.
We present all deviations from the Fermi pressure and noninteracting part of the quantum Bohm potential $T_{0s}^{\alpha\beta}$
are presented in terms of the pressure $p_{s}^{\alpha\beta}$.

The partial pressure $p_{s}^{\alpha\beta}$ is an independent function.
Therefore, equation of pressure is derived:
\begin{equation} \label{QPhrt20 eq evolution pressure}
\partial_{t}p_{s}^{\alpha\beta} +\partial_{\gamma}(v_{s}^{\gamma}p_{s}^{\alpha\beta})
+p_{s}^{\alpha\gamma}\partial_{\gamma}v_{s}^{\beta}
+p_{s}^{\beta\gamma}\partial_{\gamma}v_{s}^{\alpha}
+\partial_{\gamma}Q_{s}^{\alpha\beta\gamma}=0.  \end{equation}
Purely quantum terms like $T_{s}^{\alpha\beta}$
and the third rank quantum Bohm potential tensor $T_{s}^{\alpha\beta\gamma}$
cancel each other in equation (\ref{QPhrt20 eq evolution pressure}).

The pressure evolution equation (\ref{QPhrt20 eq evolution pressure}) obviously contains an independent function $Q_{s}^{\alpha\beta\gamma}$,
which is the third rank tensor.
Next, I derive the evolution equation for this third rank tensor:
$$\partial_{t}Q_{s}^{\alpha\beta\gamma} +\partial_{\delta}(v_{s}^{\delta}Q_{s}^{\alpha\beta\gamma})
+Q_{s}^{\alpha\gamma\delta}\partial_{\delta}v_{s}^{\beta}
+Q_{s}^{\beta\gamma\delta}\partial_{\delta}v_{s}^{\alpha}$$
\begin{equation} \label{QPhrt20 eq evolution Q}
+Q^{\alpha\beta\delta}\partial_{\delta}v_{s}^{\gamma}
+\partial_{\delta}P_{s}^{\alpha\beta\gamma\delta}
=\frac{\hbar^{2}}{4m^{3}} q_{e}n_{s}\partial^{\alpha}\partial^{\beta}\partial^{\gamma}\Phi. \end{equation}
The complete expression for this equation is presented and discussed in the Supplementary Material before truncation is made.

Equation (\ref{QPhrt20 eq evolution Q}) contains some independent functions.
Derivation of equation (\ref{QPhrt20 eq evolution Q}) is motivated by an attempt to find new quantum effects,
which are presented mainly by the first term on the right-hand side of equation (\ref{QPhrt20 eq evolution Q}).
It is proportional to $\hbar^{2}$
while other terms are proportional to $\hbar^{4}$.
However, terms proportional to $\hbar^{4}$ can be crucial in the regime of the separate spin evolution
as it is demonstrated below.
It is expected that
further derivation of hydrodynamic equations for the higher rank tensors gives small corrections
which are proportional to $\hbar^{4}$ and $\hbar^{6}$.
For instance, the derivation shows that
the evolution equation for the fifth rank tensor  contains the interaction term proportional to
$\hbar^{4}\partial^{\alpha}\partial^{\beta}\partial^{\gamma}\partial^{\delta}\partial^{\mu}\Phi_{e}$.
Therefore, I make truncation in the third rank tensor evolution equation.
To get closed set of hydrodynamic equations.
I derive equation of state for $P_{s}^{\alpha\beta\gamma\delta}$.
For the zero temperature Fermi distribution function
it has the following form:
\begin{equation} \label{QPhrt20 P s 4 equation of state} P_{s}^{\alpha\beta\gamma\delta}=\frac{6\pi^{2}}{35} I_{0}^{\alpha\beta\gamma\delta}(6\pi^{2})^{\frac{1}{3}}\frac{\hbar^{4}n_{s}^{\frac{7}{3}}}{m^{4}},\end{equation}
where
\begin{equation} \label{QPhrt20 I 4} I_{0}^{\alpha\beta\gamma\delta}=\delta^{\alpha\beta}\delta^{\gamma\delta} +\delta^{\alpha\gamma}\delta^{\beta\delta}+\delta^{\alpha\delta}\delta^{\beta\gamma}. \end{equation}
Calculation of this equation is given in the Supplementary Material.

The application of equation of state for the pressure perturbations in the Euler equation gives shift of the speed of sound.
Similarly, the application of expression (\ref{QPhrt20 P s 4 equation of state}) for the perturbations of $ P_{s}^{\alpha\beta\gamma\delta}$
allows to make an estimation of corresponding effects, but
it does not give correct coefficient.

\section{Collective excitations}

\subsection{Collective excitation in the electron gas described as the single fluid}

Consider the propagation of plane longitudinal waves in the isotropic macroscopically motionless electron-ion plasma medium.

Equilibrium concentration of electrons $n_{0e}$ is equal to the equilibrium concentration of ions $n_{0i}$.
The ions are assumed to be motionless for the consideration of the high frequency excitations.
The equilibrium velocity field are equal to zero $\textbf{v}_{0e}=0$.
The equilibrium pressure $p_{0e}^{\alpha\beta}$ is given by the isotropic Fermi pressure
$p_{0e}^{\alpha\beta}=\delta^{\alpha\beta}\cdot p_{Fe}$,
where $p_{Fe}=(3\pi^{2})^{2/3}n_{0e}^{5/3}\hbar^{2}/5m_{e}$,
$p_{0e}^{\alpha\beta}=p_{0u}^{\alpha\beta}+p_{0d}^{\alpha\beta}$ is the superposition of partial pressures.
The equilibrium third rank tensor is equal to zero for the zero temperature isotropic fermions $Q_{0e}^{\alpha\beta\gamma}=0$
(see Supplementary Material).
The equilibrium fourth rank tensor is expressed via the equilibrium concentration is accordance with the expression (\ref{QPhrt20 P s 4 equation of state}):
\begin{equation} \label{QPhrt20 P s 4 equation of state Equilibrium} P_{0e}^{\alpha\beta\gamma\delta}=P_{0u}^{\alpha\beta\gamma\delta}+P_{0d}^{\alpha\beta\gamma\delta}
=\frac{3\pi^{2}}{35} I_{0}^{\alpha\beta\gamma\delta}(3\pi^{2})^{\frac{1}{3}}\frac{\hbar^{4}n_{0e}^{\frac{7}{3}}}{m^{4}}.\end{equation}
The scalar potential of the electric field is equal to zero in the equilibrium state $\Phi_{0}=0$.

Let me consider the small amplitude perturbations of all material fields involved in the presented model
for perturbations propagating parallel to $x$-direction:
$n_{e}=n_{0e}+\delta n_{e}$,
$v_{e}^{x}=0+\delta v_{e}^{x}$,
$\delta p_{e}^{\alpha\beta}=p_{0e}^{\alpha\beta} +\delta^{x\alpha}\delta^{x\beta}\delta p_{e}^{xx}$,
$Q_{e}^{\alpha\beta\gamma}=0+\delta^{x\alpha}\delta^{x\beta}\delta^{x\gamma}\delta Q_{e}^{xxx}$,
$P_{e}^{\alpha\beta\gamma\delta}=P_{0e}^{\alpha\beta\gamma\delta}+\delta^{x\alpha}\delta^{x\beta}\delta^{x\gamma}\delta^{x\delta}\delta P_{e}^{xxxx}$,
and
$\Phi=0+\delta\Phi$,
where
\begin{equation} \label{QPhrt20 P e xxxx perturb}\delta P_{e}^{xxxx}=
\frac{(3\pi^{2})^{4/3}}{5}\frac{\hbar^{4}n_{0e}^{4/3}}{m_{e}^{4}}\delta n_{e}
=\frac{1}{5}v_{Fe}^{4}\delta n_{e},\end{equation}
with the Fermi velocity $v_{Fe}=(3\pi^{2}n_{0e})^{1/3}\hbar/m_{e}$,
and $I_{0}^{xxxx}=3$.

Perturbation of each function is presented in the form of plane wave:
\begin{equation} \label{QPhrt20} \left(
                                   \begin{array}{c}
                                     \delta n_{e} \\
                                     \delta v_{e}^{x} \\
                                     \delta p_{e}^{xx} \\
                                     \delta Q_{e}^{xxx} \\
                                     \delta\Phi \\
                                   \end{array}
                                 \right)=\left(
                                  \begin{array}{c}
                                     N_{e} \\
                                     V_{e} \\
                                     P_{e} \\
                                     Q_{e} \\
                                     \Phi_{ampl} \\
                                   \end{array}
                                 \right) e^{-\imath\omega t+\imath k x}.
\end{equation}

The described procedure leads to the standard form of the linearized continuity equation:
\begin{equation} \label{QPhrt20 cont lin} \omega\delta n_{e}=kn_{0e}\delta v^{x}_{e}. \end{equation}

The linearized Euler equation has one modification.
The single element of the pressure tensor $\delta p_{e}^{xx}$ is presented as the independent function:
\begin{equation} \label{QPhrt20 Euler lin} \omega m_{e}n_{0e}\delta v^{x}_{e}-k \delta p_{e}^{xx} -\frac{\hbar^{2}k^{3}}{4m_{e}}\delta n_{e}
=q_{e}n_{0e}k\delta\Phi.\end{equation}

The perturbation of the pressure $\delta p_{e}^{xx}$ can be found from the linearized pressure evolution equation:
\begin{equation} \label{QPhrt20 pressure evol lin} \omega \delta p_{e}^{xx}=3kp_{0e}^{xx}\delta v_{e}^{x}+k\delta Q_{e}^{xxx}.\end{equation}
The first term on the right-hand side of the pressure evolution equation (\ref{QPhrt20 pressure evol lin}) provides the partial expression for the pressure perturbation in accordance with the kinetic model \cite{Landau Vol X}, \cite{Aleksandrov Rukhadze Book}, \cite{Andreev PoP 16 sep kin}:
\begin{equation} \label{QPhrt20} \delta p^{xx}_{partial}=3\frac{p_{0e}^{xx}}{n_{0e}}\delta n_{e}=\frac{3}{5}v_{Fe}^{2}\delta n_{e}. \end{equation}
Novel quantum effects enters the model via the second term on the right-hand side of equation (\ref{QPhrt20 pressure evol lin}).

The linearized equation for the third order rank tensor evolution has the following form:
\begin{equation} \label{QPhrt20 Q evol lin}
\omega \delta Q_{e}^{xxx}=k\delta P_{e}^{xxxx}
+\frac{\hbar^{2}}{4m^{2}} q_{e}n_{0e}k^{3}\delta\Phi. \end{equation}

The presented model contains the traditional form of the linearized Poisson equation:
\begin{equation} \label{QPhrt20 PE lin} k^{2}\delta\Phi=4\pi q_{e}\delta n_{e}. \end{equation}


Linearized equations (\ref{QPhrt20 cont lin})-(\ref{QPhrt20 Q evol lin}) lead to the following dispersion equation,
which is the quadratic equation relatively $\omega^{2}$:
$$\omega^{4}-\omega^{2}\biggl(\omega_{Le}^{2}+\frac{3p_{0}^{xx}}{mn_{0}}k^{2}+\frac{\hbar^{2}k^{4}}{4m^{2}}\biggr)$$
\begin{equation} \label{QPhrt20 Disp eq SF}
-\biggl(\omega_{Le}^{2}\frac{\hbar^{2}k^{4}}{4m^{2}}+k^{4}\frac{\delta P_{0}^{xxxx}}{\delta n_{0}}\biggr)=0, \end{equation}
where
$\omega_{Le}^{2}=4\pi e^{2}n_{0e}/m_{e}$ is the Langmuir frequency.
The explicit form of the last term in equation (\ref{QPhrt20 Disp eq SF}) can be found from equation (\ref{QPhrt20 P e xxxx perturb}).
However, it is kept in unexplicit form to track its source.

I believe it is useful to specify that
I consider the regime for the Langmuir waves.
However, the quantum effects entering the model via the evolution of the third rank tensor gives additional solution.

Here, I present the solution of equation (\ref{QPhrt20 Disp eq SF}):
$$\omega_{\pm}^{2}=\frac{1}{2}\biggl\{\omega_{Le}^{2}+\frac{3p_{0}^{xx}}{mn_{0}}k^{2}+\frac{\hbar^{2}k^{4}}{4m^{2}}$$
$$\pm\biggl[\biggl(\omega_{Le}^{2}+\frac{3p_{0}^{xx}}{mn_{0}}k^{2}+\frac{\hbar^{2}k^{4}}{4m^{2}}\biggr)^{2}$$
\begin{equation} \label{QPhrt20 sol of DE SF}
+4\omega_{Le}^{2}\frac{\hbar^{2}k^{4}}{4m^{2}}+k^{4}\frac{\delta P_{0}^{xxxx}}{\delta n_{0}}\biggr]^{1/2}\biggr\}, \end{equation}
where the terms proportional to $\hbar^{4}$ are dropped.

Solution $\omega_{-}^{2}$ can be rewritten in the following form
\begin{equation} \label{QPhrt20 omega - unexplicit} \omega_{-}^{2}= \frac{\omega_{-}^{2}\omega_{+}^{2}}{\omega_{+}^{2}} =-\frac{\omega_{Le}^{2}\frac{\hbar^{2}k^{4}}{4m^{2}}+\frac{1}{5}v_{Fe}^{4}k^{4}}{\omega_{+}^{2}}.\end{equation}

It is clear that the second solution exists in the quasi-classic limit.
However, it is suppressed by the small thermal velocity $\delta P_{0}^{xxxx}/\delta n_{0}\sim v_{T}^{4}$ in the small temperature limit.
In the quantum regime with temperature $T$ below the Fermi temperature $T_{Fe}=mv_{Fe}^{2}/2$
the quasi-classic contribution is also suppressed since it is proportional to the high degree of the Planck constant $\hbar^{4}$.
While the quantum-interaction part demonstrated in this paper is proportional to the $\hbar^{2}$ and the large frequency value $\omega_{Le}^{2}$.
Therefore, existence of solution mainly caused by the first term on the right-hand side of equation (\ref{QPhrt20 omega - unexplicit})
found from the right-hand side of equation (\ref{QPhrt20 eq evolution Q}).

Let me consider the small wave vector limit of obtained solution:
\begin{equation} \label{QPhrt20} \omega_{+}^{2}=\omega_{Le}^{2}, \end{equation}
and
\begin{equation} \label{QPhrt20} \omega_{-}^{2}=-\frac{\hbar^{2}k^{4}}{4m^{2}}-\frac{1}{5}k^{4}v_{Fe}^{2}r_{DF}^{2}, \end{equation}
where $r_{DF}=v_{Fe}/\omega_{Le}$ is the Debye radius for the degenerate electrons.

\subsection{Spin-electron-acoustic waves}

Consider the propagation of plane longitudinal waves in the direction of the external magnetic field.
The external magnetic field is one of mechanisms of the spin polarization formation.
Magnetic conductive materials create the spontaneous spin polarization of the lattice and the electron gas.
For these materials spin polarization of electrons is nonzero even for the zero external magnetic field.
The z-axis is directed parallel to the equilibrium spin polarization $S_{0z}=n_{0\uparrow}-n_{0\downarrow}$.

The structure of the equilibrium state and the form of the perturbations are similar to the described above for the single fluid regime.
Evolution of perturbations leads to the following dispersion equation:
\begin{equation} \label{QPhrt20 DE SSE}
1=\sum_{s}\frac{\omega_{Ls}^{2}(1+\frac{\hbar^{2}k^{4}}{4m^{2}\omega^{2}})}{ \omega^{2}-\frac{3p_{0s}^{xx}}{mn_{0s}}k^{2}-\frac{\hbar^{2}k^{4}}{4m^{2}}-\frac{k^{4}}{\omega^{2}}\frac{\delta P_{0s}^{xxxx}}{\delta n_{0s}}}, \end{equation}
where $\omega_{Ls}^{2}=4\pi e^{2}n_{0s}/m$ is the partial Langmuir frequency.

Dispersion equation (\ref{QPhrt20 DE SSE}) is obtained for the regime,
where two waves (the Langmuir wave and the spin-electron-acoustic wave \cite{Andreev PRE 15 SEAW}) exist
in the traditional separate-spin-evolution quantum hydrodynamics construct of two continuity equations and two Euler equations \cite{Andreev PRE 15 SEAW}.
Here, I find four waves.
Hence, there are two new solutions.
One of them is found in the single fluid regime.
Therefore, the regime of the separate spin evolution brings the novel solution.

If I drop the contribution of the third rank tensor,
equation (\ref{QPhrt20 DE SSE}) simplifies to
\begin{equation} \label{QPhrt20 DE SSE no Q}
1=\sum_{s}\frac{\omega_{Ls}^{2}}{ \omega^{2}-\frac{3p_{0s}^{xx}}{mn_{0s}}k^{2}-\frac{\hbar^{2}k^{4}}{4m^{2}}}. \end{equation}
Equation (\ref{QPhrt20 DE SSE no Q}) includes the contribution of the pressure perturbations from the pressure evolution equation.
Therefore, the speed of sound for the spin-electron-acoustic waves corresponds to the kinetic model \cite{Andreev PoP 16 sep kin}
in contrast with hydrodynamics based on the continuity and Euler equations \cite{Andreev PRE 15 SEAW}.

\section{Conclusion}

The extended hydrodynamic model demonstrating novel quantum effects has been developed.
These quantum effects appear in addition to the well-known quantum Bohm potential and spin effects.
The model has been presented for the electrostatic regime.
Hence, the Coulomb interaction is considered.
Therefore, the quantum corrections to the Coulomb interaction is found via evolution of the third rank tensor.
However, if one includes the spin-spin interaction
this model gives the quantum part for the spin-spin interaction, or any other interaction.
Such generalizations will be considered in the future publications.
Here, novel quantum phenomena in plasmas are demonstrated on the simple examples.
So, other phenomena do not cover the found quantum effects.
Presented model gives the background for re-innovation of quantum phenomena caused by the quantum Bohm potential.

\section{Acknowledgements}
Work is supported by the Russian Foundation for Basic Research (grant no. 20-02-00476).

\newpage

\section{SUPPLEMENTARY MATERIAL}

\subsection{Basic definitions, microscopic Hamiltonian of the system, and the general structure of the continuity equation}

Description of collective behavior can be started with the concentration or the number of particles
\begin{equation}\label{QPhrt20 concentration def} n=\int
dR\sum_{i=1}^{N}\delta(\textbf{r}-\textbf{r}_{i})\Psi^{\dag}(R,t)\Psi(R,t),\end{equation}
which is the first collective variable in our model.
Other collective variables appear during the derivation.
Equation (\ref{QPhrt20 concentration def}) contains the following notations
$dR=\prod_{i=1}^{N}d\textbf{r}_{i}$ is the element of volume in $3N$ dimensional configurational space,
with $N$ is the number of electrons.
Symbol $^{\dag}$ means the Hermitian conjugation.

If we consider the separate spin evolution hydrodynamics
we need to split concentration of electrons
on two parts: $n_{e}=n_{\uparrow}+n_{\downarrow}$.
This separation is made in accordance with the structure of the wave function
\begin{equation}\label{QPhrt20 concentration def partial} n_{s}=\int
dR\sum_{i=1}^{N}\delta(\textbf{r}-\textbf{r}_{i})\Psi_{s}^{*}(R,t)\Psi_{s}(R,t),\end{equation}
where
\begin{equation}\label{QPhrt20 WF} \Psi(R,t)=\left(
                                                                \begin{array}{c}
                                                                  \Psi_{\uparrow}(R,t) \\
                                                                  \Psi_{\downarrow}(R,t) \\
                                                                \end{array}
                                                              \right)
.\end{equation}
Symbol $^{*}$ means the complex conjugation.
Sum in equation (\ref{QPhrt20 concentration def partial}) is made for all electrons.
The probability to have a specified spin projection is kept in the wave function.
Spin polarization of each electron can be partial.


\begin{equation}\label{QPhrt20 Hamiltonian micro}
\hat{H}=\sum_{i=1}^{N}\biggl(\frac{\hat{\textbf{p}}^{2}_{i}}{2m_{i}}\biggr)
+\frac{1}{2}\sum_{i,j\neq i}\frac{q_{e}^{2}}{\mid \textbf{r}_{i}-\textbf{r}_{j}\mid} ,\end{equation}
where $m_{i}$ is the mass of i-th particle,
$\hat{\textbf{p}}_{i}=-\imath\hbar\nabla_{i}$ is the momentum of i-th particle.

$$\textbf{j}_{s}
=\int dR\sum_{i=1}^{N}\delta(\textbf{r}-\textbf{r}_{i})\times$$
\begin{equation}\label{QPhrt20 j def}
\times\frac{1}{2m_{i}}(\Psi_{s}^{*}(R,t)\hat{\textbf{p}}_{i}\Psi_{s}(R,t)+c.c.),\end{equation}
with $c.c.$ is the complex conjugation.

\subsection{General structure of the momentum balance equation}

Definition of current (\ref{QPhrt20 j def}) allows to derive the Euler equation for the current (momentum density) evolution
\begin{equation} \label{QPhrt20 Euler eq via j}
\partial_{t}j_{s}^{\alpha}+\partial_{\beta}\Pi_{s}^{\alpha\beta}
=\frac{1}{m}F^{\alpha}_{int}, \end{equation}
where
$$\Pi_{s}^{\alpha\beta}=\int dR\sum_{i=1}^{N}\delta(\textbf{r}-\textbf{r}_{i}) \frac{1}{4m^{2}}
[\Psi_{s}^{*}(R,t)\hat{p}_{i}^{\alpha}\hat{p}_{i}^{\beta}\Psi_{s}(R,t)$$
\begin{equation} \label{QPhrt20 Pi def} +\hat{p}_{i}^{\alpha *}\Psi_{s}^{*}(R,t)\hat{p}_{i}^{\beta}\Psi_{s}(R,t)+c.c.] \end{equation}
is the momentum flux,
and
\begin{equation} \label{QPhrt20 F alpha def via n2}
F^{\alpha}_{int}=-\int (\partial^{\alpha}U(\textbf{r}-\textbf{r}'))
n_{2,ss'}(\textbf{r},\textbf{r}',t)d\textbf{r}', \end{equation}
with the two-particle concentration
$$n_{2}(\textbf{r},\textbf{r}',t)$$
\begin{equation} \label{QPhrt20 n2 def} =\int
dR\sum_{i,j=1,j\neq i}^{N}\delta(\textbf{r}-\textbf{r}_{i})\delta(\textbf{r}'-\textbf{r}_{j})\Psi_{s}^{*}(R,t)\Psi_{s}(R,t) ,\end{equation}
and the Coulomb interaction potential
\begin{equation} \label{QPhrt20} U(\textbf{r}-\textbf{r}')=\frac{q_{e}^{2}}{\mid \textbf{r}-\textbf{r}'\mid}. \end{equation}

The Euler equation has simple structure.
It shows that the time evolution of the current or the momentum density $\textbf{j}$ is caused by the mechanisms.
One of them is the kinetic momentum flux presented in the left-hand side.
It is related to the motion of particles being in the fixed states.
The second mechanism is the interaction.
Same structure repeats itself in other hydrodynamic equations for the physical quantities with the higher rank tensors.
The evolution of the chosen quantities caused by its flux and due to the interaction.

\subsection{General structure of equation for the second order tensor}

Extending the set of hydrodynamic equations we can derive the equation for the momentum flux evolution.
Consider the time evolution of the momentum flux (\ref{QPhrt20 Pi def}) using the Schrodinger equation with Hamiltonian (\ref{QPhrt20 Hamiltonian micro})
and derive the momentum flux evolution equation
\begin{equation} \label{QPhrt20 eq for Pi alpha beta}
\partial_{t}\Pi_{s}^{\alpha\beta}+\partial_{\gamma}M_{s}^{\alpha\beta\gamma}
=\frac{1}{m}(F^{\alpha\beta}+F^{\beta\alpha}), \end{equation}
where
\begin{equation} \label{QPhrt20 F alpha beta def} F^{\alpha\beta}=-\int[\partial^{\alpha}U(\textbf{r}-\textbf{r}')]j_{2,ss'}^{\beta}(\textbf{r},\textbf{r}',t)d\textbf{r}'\end{equation}
represents the interaction,
$$M_{s}^{\alpha\beta\gamma}=\int dR\sum_{i=1}^{N}\delta(\textbf{r}-\textbf{r}_{i}) \frac{1}{8m_{i}^{3}}\biggl[\Psi_{s}^{*}(R,t)\hat{p}_{i}^{\alpha}\hat{p}_{i}^{\beta}\hat{p}_{i}^{\gamma}\Psi_{s}(R,t)$$
$$+\hat{p}_{i}^{\alpha *}\Psi_{s}^{*}(R,t)\hat{p}_{i}^{\beta}\hat{p}_{i}^{\gamma}\Psi_{s}(R,t)
+\hat{p}_{i}^{\alpha *}\hat{p}_{i}^{\gamma *}\Psi_{s}^{*}(R,t)\hat{p}_{i}^{\beta}\Psi_{s}(R,t)$$
\begin{equation} \label{QPhrt20 M alpha beta gamma def}
+\hat{p}_{i}^{\gamma *}\Psi_{s}^{*}(R,t)\hat{p}_{i}^{\alpha}\hat{p}_{i}^{\beta}\Psi_{s}(R,t)+c.c.\biggr] \end{equation}
is the flux of the momentum flux,
and
$$\textbf{j}_{2,ss'}(\textbf{r},\textbf{r}',t)=\int
dR\sum_{i,j\neq i}\delta(\textbf{r}-\textbf{r}_{i})\delta(\textbf{r}'-\textbf{r}_{j})\times$$
\begin{equation} \label{QPhrt20 j 2 def}
\times\frac{1}{2m_{i}}(\Psi_{s}^{*}(R,t)\hat{\textbf{p}}_{i}\Psi_{s}(R,t)+c.c.) .\end{equation}
If quantum correlations are dropped function $j_{2}^{\alpha}(\textbf{r},\textbf{r}',t)$
splits on product of the current $j^{\alpha}(\textbf{r},t)$ and the concentration $n(\textbf{r}',t)$.

\subsection{General structure of equation for the third order tensor}

General structure of the evolution equation for the third rank tensor $M_{s}^{\alpha\beta\gamma}$:
$$\partial_{t}M_{s}^{\alpha\beta\gamma}+\partial_{\delta}R_{s}^{\alpha\beta\gamma\delta}$$
\begin{equation} \label{QPhrt20 eq for M alpha beta gamma}
=\frac{1}{m}(F_{Q}^{\alpha\beta\gamma}
+F^{\beta\alpha\gamma}+F^{\beta\alpha\gamma}+F^{\beta\alpha\gamma}), \end{equation}
where
\begin{equation} \label{QPhrt20 F alpha beta gamma def} F_{Q}^{\alpha\beta\gamma}=\frac{\hbar^{2}}{4m^{3}}
\int[\partial^{\alpha}\partial^{\beta}\partial^{\gamma}U(\textbf{r}-\textbf{r}')]n_{2}(\textbf{r},\textbf{r}',t)d\textbf{r}'\end{equation}
is the quantum part of interaction reported in this paper,
\begin{equation} \label{QPhrt20 F alpha beta gamma def} F^{\alpha\beta\gamma}=-\int[\partial^{\alpha}U(\textbf{r}-\textbf{r}')]\Pi_{2}^{\beta\gamma}(\textbf{r},\textbf{r}',t)d\textbf{r}'\end{equation}
is the quasi-classic part of interaction,
$$R_{s}^{\alpha\beta\gamma\delta}=\int dR\sum_{i=1}^{N}\delta(\textbf{r}-\textbf{r}_{i}) \frac{1}{16m_{i}^{4}}\biggl[\Psi_{s}^{*}(R,t)\hat{p}_{i}^{\alpha}\hat{p}_{i}^{\beta}\hat{p}_{i}^{\gamma}\hat{p}_{i}^{\delta}\Psi_{s}(R,t)$$
$$+\hat{p}_{i}^{\alpha *}\Psi_{s}^{*}(R,t)\hat{p}_{i}^{\beta}\hat{p}_{i}^{\gamma}\hat{p}_{i}^{\delta}\Psi_{s}(R,t)
+\hat{p}_{i}^{\alpha *}\hat{p}_{i}^{\gamma *}\hat{p}_{i}^{\delta *}\Psi_{s}^{*}(R,t)\hat{p}_{i}^{\beta}\Psi_{s}(R,t)$$

$$+\hat{p}_{i}^{\gamma *}\Psi_{s}^{*}(R,t)\hat{p}_{i}^{\alpha}\hat{p}_{i}^{\beta}\hat{p}_{i}^{\delta}\Psi_{s}(R,t)
+\hat{p}_{i}^{\alpha *}\hat{p}_{i}^{\beta *}\hat{p}_{i}^{\gamma *}\Psi_{s}^{*}(R,t)\hat{p}_{i}^{\delta}\Psi_{s}(R,t)$$

$$+\hat{p}_{i}^{\alpha *}\hat{p}_{i}^{\delta *}\Psi_{s}^{*}(R,t)\hat{p}_{i}^{\beta}\hat{p}_{i}^{\gamma}\Psi_{s}(R,t)
+\hat{p}_{i}^{\alpha *}\hat{p}_{i}^{\gamma *}\Psi_{s}^{*}(R,t)\hat{p}_{i}^{\beta}\hat{p}_{i}^{\delta}\Psi_{s}(R,t)$$
\begin{equation} \label{QPhrt20 P alpha beta gamma delta def}
+\hat{p}_{i}^{\gamma *}\hat{p}_{i}^{\delta *}\Psi_{s}^{*}(R,t)\hat{p}_{i}^{\alpha}\hat{p}_{i}^{\beta}\Psi_{s}(R,t)+c.c.\biggr] \end{equation}
is the flux of $M_{s}^{\alpha\beta\gamma}$
and
$$\Pi_{2,ss'}^{\alpha\beta}(\textbf{r},\textbf{r}',t)=\int
dR\sum_{i,j\neq i}\delta(\textbf{r}-\textbf{r}_{i})
\delta(\textbf{r}'-\textbf{r}_{j})\frac{1}{4m_{i}^{2}}\times$$
\begin{equation} \label{QPhrt20 Pi 2 def}
\times(\Psi_{s}^{*}(R,t)\hat{p}_{i}^{\alpha}\hat{p}_{i}^{\beta}\Psi_{s}(R,t)
+(\hat{p}_{i}^{\alpha}\Psi_{s}(R,t))^{*}\hat{p}_{i}^{\beta}\Psi_{s}(R,t)+c.c.) .\end{equation}
If quantum correlations are dropped function $\Pi_{2}^{\alpha\beta}(\textbf{r},\textbf{r}',t)$
splits on product of the momentum flux $\Pi^{\alpha\beta}(\textbf{r},t)$ and the concentration $n(\textbf{r}',t)$.

General untruncated form of the equation for the "thermal" part of the third rank tensor (the part defined in comoving frame) has the following form
$$\partial_{t}\tilde{Q}_{s}^{\alpha\beta\gamma} +\partial_{\delta}(v_{s}^{\delta}\tilde{Q}_{s}^{\alpha\beta\gamma})
+\tilde{Q}_{s}^{\alpha\gamma\delta}\partial_{\delta}v_{s}^{\beta}
+\tilde{Q}_{s}^{\beta\gamma\delta}\partial_{\delta}v_{s}^{\alpha}$$
$$+\tilde{Q}_{s}^{\alpha\beta\delta}\partial_{\delta}v_{s}^{\gamma}
+\partial_{\delta}(P_{s}^{\alpha\beta\gamma\delta}+T_{s}^{\alpha\beta\gamma\delta})
=\frac{\hbar^{2}}{4m^{3}} q_{e}n_{s}\partial^{\alpha}\partial^{\beta}\partial^{\gamma}\Phi$$
$$+\frac{1}{mn}[(p_{s}^{\alpha\beta}+T_{s}^{\alpha\beta})\partial^{\delta}(p_{s}^{\gamma\delta}+T_{s}^{\gamma\delta})
+(p_{s}^{\alpha\gamma}+T_{s}^{\alpha\gamma})\times$$
\begin{equation} \label{QPhrt20 SM eq evolution Q}
\times\partial^{\delta}(p_{s}^{\beta\delta}+T_{s}^{\beta\delta})
+(p_{s}^{\beta\gamma}+T_{s}^{\beta\gamma})\partial^{\delta}(p_{s}^{\alpha\delta}+T_{s}^{\alpha\delta})],  \end{equation}
where
$\tilde{Q}_{s}^{\alpha\beta\gamma}=Q_{s}^{\alpha\beta\gamma}+T_{s}^{\alpha\beta\gamma}$,
\begin{equation} \label{QPhrt20 SM T 3 expl}T_{s}^{\alpha\beta\gamma}
=-\frac{\hbar^{2}}{12m^{2}}n_{s}(\partial^{\alpha}\partial^{\beta} v_{s}^{\gamma}
+\partial^{\alpha}\partial^{\gamma} v_{s}^{\beta}
+\partial^{\beta}\partial^{\gamma} v_{s}^{\alpha})\end{equation}
is the third rank tensor analog of the quantum Bohm potential,
the fourth rank tensor $P_{s}^{\alpha\beta\gamma\delta}$ is constructed on the thermal velocities or the velocities in the local frame,
basically the fourth rank tensor $P_{s}^{\alpha\beta\gamma\delta}$ is the analog of the pressure tensor with higher tensor rank,
$$T_{s,lin}^{\alpha\beta\gamma\delta}
=\frac{\hbar^{4}}{8m^{4}}
\sqrt{n_{s}}\partial^{\alpha}\partial^{\beta}\partial^{\gamma}\partial^{\delta}\sqrt{n_{s}}$$
$$+2p_{s}^{\alpha\beta}T_{s}^{\gamma\delta}/n_{s}
+2p_{s}^{\alpha\gamma}T_{s}^{\beta\delta}/n_{s} +2p_{s}^{\alpha\delta}T_{s}^{\beta\gamma}/n_{s}$$
\begin{equation} \label{QPhrt20 SM T 4 expl short} +2p_{s}^{\beta\gamma}T_{s}^{\alpha\delta}/n_{s}
+2p_{s}^{\beta\delta}T_{s}^{\alpha\gamma}/n_{s} +2p_{s}^{\gamma\delta}T_{s}^{\alpha\beta}/n_{s}.\end{equation}
is the main part of the fourth rank tensor analog of the quantum Bohm potential.

I should omit the term proportional to the spatial derivatives of the fourth and second rank kinetic tensors
in accordance with estimations presented in Ref. \cite{Tokatly PRB 00},
but I keep $\partial_{\delta}P_{s}^{\alpha\beta\gamma\delta}$ to get some
rough estimations of the fourth rank pressure-like tensor contribution.

\subsection{Equilibrium expressions for the pressure and pressure-like third and fourth rank tensors}

Perturbations of pressure tensor and the third rank tensor can be found from the corresponding equations of evolution.
However, their equilibrium values are found via the equilibrium distribution function
chosen in the form of the Fermi step function:
\begin{equation} \label{QPhrt20}  p_{0s}^{\alpha\beta}=m\int_{0}^{p_{Fs}} p^{\alpha}p^{\beta} \frac{d^{3}p}{(2\pi\hbar)^{3}}, \end{equation}
and
\begin{equation} \label{QPhrt20}  Q_{0s}^{\alpha\beta\gamma}=\int_{0}^{p_{Fs}} p^{\alpha}p^{\beta}p^{\gamma} \frac{d^{3}p}{(2\pi\hbar)^{3}}=0. \end{equation}

The equation of state for the thermal part (or the Pauli blocking part) of the fourth rank tensor is also found via the equilibrium distribution function
chosen in the form of the Fermi step function:
\begin{equation} \label{QPhrt20}  P_{0s}^{\alpha\beta\gamma\delta}=\int_{0}^{p_{Fs}} p^{\alpha}p^{\beta}p^{\gamma}p^{\delta} \frac{d^{3}p}{(2\pi\hbar)^{3}}, \end{equation}
where symbol $p$ with no indexes is the momentum, $p_{Fs}=(6\pi^{2}n_{0s})^{1/3}\hbar$ is the partial Fermi momentum.

\subsection{Fourth rank quantum Bohm potential}

The fourth rank tensor,
which is similar in nature with the quantum Bohm potential,
appears in the equation for evolution of the third rank tensor (\ref{QPhrt20 eq evolution Q}).
It has rather complex form.
Therefore,
it is not demonstrated in the main part of the paper.

Hence, this tensor is given as the superposition of three parts:
\begin{equation} \label{QPhrt20 T 4 expl}
T_{s}^{\alpha\beta\gamma\delta}=\sum_{i=1}^{3}T_{si}^{\alpha\beta\gamma\delta}.\end{equation}

The first part can be approximately written via the concentration of fermions:
$$T_{s1}^{\alpha\beta\gamma\delta}
=\frac{\hbar^{4}}{8m^{4}}
\biggl[\sqrt{n}\partial^{\alpha}\partial^{\beta}\partial^{\gamma}\partial^{\delta}\sqrt{n}$$
$$+\partial^{\alpha}\partial^{\beta}\sqrt{n}\cdot\partial^{\gamma}\partial^{\delta}\sqrt{n}
+\partial^{\alpha}\partial^{\gamma}\sqrt{n}\cdot\partial^{\beta}\partial^{\delta}\sqrt{n}
+\partial^{\alpha}\partial^{\delta}\sqrt{n}\cdot\partial^{\beta}\partial^{\gamma}\sqrt{n}$$
$$-\partial^{\alpha}\sqrt{n}\cdot\partial^{\beta}\partial^{\gamma}\partial^{\delta}\sqrt{n}
-\partial^{\beta}\sqrt{n}\cdot\partial^{\alpha}\partial^{\gamma}\partial^{\delta}\sqrt{n}$$
\begin{equation} \label{QPhrt20 T 4 expl p1}
-\partial^{\gamma}\sqrt{n}\cdot\partial^{\alpha}\partial^{\beta}\partial^{\delta}\sqrt{n}
-\partial^{\delta}\sqrt{n}\cdot\partial^{\alpha}\partial^{\beta}\partial^{\gamma}\sqrt{n}\biggr]. \end{equation}
Similar approximation is used for the quantum Bohm potential in equation (\ref{QPhrt20 Bohm tensor single part}).

The second part of tensor $T_{s}^{\alpha\beta\gamma\delta}$ contains the traditional quantum Bohm potential:
$$T_{s2}^{\alpha\beta\gamma\delta}=
2\biggl[v^{\gamma}v^{\delta}T_{micro}^{\alpha\beta}
+v^{\gamma}\langle u^{\delta}t^{\alpha\beta}\rangle
+v^{\delta}\langle u^{\gamma}t^{\alpha\beta}\rangle
+\langle u^{\gamma}u^{\delta}t^{\alpha\beta}\rangle \biggr]$$

$$+2\biggl[v^{\beta}v^{\delta}T_{micro}^{\alpha\gamma}
+v^{\beta}\langle u^{\delta}t^{\alpha\gamma}\rangle
+v^{\delta}\langle u^{\beta}t^{\alpha\gamma}\rangle
+\langle u^{\beta}u^{\delta}t^{\alpha\gamma}\rangle \biggr]$$

$$+2\biggl[v^{\gamma}v^{\beta}T_{micro}^{\alpha\delta}
+v^{\gamma}\langle u^{\beta}t^{\alpha\delta}\rangle
+v^{\beta}\langle u^{\gamma}t^{\alpha\delta}\rangle
+\langle u^{\gamma}u^{\beta}t^{\alpha\delta}\rangle \biggr]$$

$$+2\biggl[v^{\alpha}v^{\delta}T_{micro}^{\beta\gamma}
+v^{\alpha}\langle u^{\delta}t^{\beta\gamma}\rangle
+v^{\delta}\langle u^{\alpha}t^{\beta\gamma}\rangle
+\langle u^{\alpha}u^{\delta}t^{\beta\gamma}\rangle \biggr]$$

$$+2\biggl[v^{\alpha}v^{\gamma}T_{micro}^{\beta\delta}
+v^{\gamma}\langle u^{\alpha}t^{\beta\delta}\rangle
+v^{\alpha}\langle u^{\gamma}t^{\beta\delta}\rangle
+\langle u^{\alpha}u^{\gamma}t^{\beta\delta}\rangle \biggr]$$

\begin{equation} \label{QPhrt20 T 4 expl p2}
+2\biggl[v^{\alpha}v^{\beta}T_{micro}^{\gamma\delta}
+v^{\alpha}\langle u^{\beta}t^{\gamma\delta}\rangle
+v^{\beta}\langle u^{\alpha}t^{\gamma\delta}\rangle
+\langle u^{\alpha}u^{\beta}t^{\gamma\delta}\rangle \biggr],\end{equation}
where
\begin{equation} \label{QPhrt20}t^{\alpha\beta}=\frac{\hbar^{2}}{4m^{2}}(a\partial^{\alpha}\partial^{\beta}a-\partial^{\alpha}a\cdot\partial^{\beta}a),\end{equation}
$$T_{micro}^{\alpha\beta}=\frac{\hbar^{2}}{4m^{2}}\int dR\sum_{i=1}^{N}\delta(\textbf{r}-\textbf{r}_{i})t^{\alpha\beta}$$
\begin{equation} \label{QPhrt20 T micro}=\int dR\sum_{i=1}^{N}\delta(\textbf{r}-\textbf{r}_{i})(a\partial^{\alpha}\partial^{\beta}a-\partial^{\alpha}a\cdot\partial^{\beta}a),\end{equation}
where
$T_{micro}^{\alpha\beta}=\langle t^{\alpha\beta}\rangle$.
Equation (\ref{QPhrt20 T micro}) approximately gives the quantum Bohm potential (\ref{QPhrt20 Bohm tensor single part})
$T_{micro}^{\alpha\beta}\approx T^{\alpha\beta}$.
However, other expressions like $\langle u^{\alpha}t^{\beta\delta}\rangle$ has no simple expression.
Using the theorem on average we can make the following approximation
$\langle u^{\alpha}t^{\beta\delta}\rangle=(T^{\alpha\beta}/n)\langle a^{2} u^{\alpha}\rangle=0$,
since $\langle a^{2} u^{\alpha}\rangle=0$ by definition of the thermal velocity.
Moreover, we find $\langle u^{\alpha}u^{\beta}t^{\gamma\delta}\rangle\approx (T^{\gamma\delta}/n)p^{\alpha\beta}$.
These expressions are used as the equation of state for described functions.

The third part of tensor $T_{s}^{\alpha\beta\gamma\delta}$ construct of the velocities
$$T_{s3}^{\alpha\beta\gamma\delta}=\frac{\hbar^{2}}{24m^{2}}
\Biggl[nv^{\alpha}[\partial^{\beta}\partial^{\gamma}v^{\delta}+\partial^{\gamma}\partial^{\delta}v^{\beta}+\partial^{\beta}\partial^{\delta}v^{\gamma}]$$
$$+nv^{\beta}[\partial^{\alpha}\partial^{\gamma}v^{\delta}+\partial^{\alpha}\partial^{\delta}v^{\gamma}+\partial^{\gamma}\partial^{\delta}v^{\alpha}]$$
$$+nv^{\gamma}[\partial^{\alpha}\partial^{\beta}v^{\delta}+\partial^{\alpha}\partial^{\delta}v^{\beta}+\partial^{\beta}\partial^{\delta}v^{\alpha}]$$
$$+nv^{\delta}[\partial^{\alpha}\partial^{\beta}v^{\gamma}+\partial^{\alpha}\partial^{\gamma}v^{\beta}+\partial^{\beta}\partial^{\gamma}v^{\alpha}]$$

$$+v^{\alpha}[\langle a^2 \partial^{\beta}\partial^{\gamma}u^{\delta}\rangle+ \langle a^2\partial^{\beta}\partial^{\delta}u^{\gamma}\rangle+ \langle a^2\partial^{\gamma}\partial^{\delta}u^{\beta}\rangle]$$
$$+v^{\beta}[\langle a^2 \partial^{\alpha}\partial^{\gamma}u^{\delta}\rangle+ \langle a^2 \partial^{\alpha}\partial^{\delta}u^{\gamma}\rangle+ \langle a^2 \partial^{\gamma}\partial^{\delta}u^{\alpha}\rangle]$$
$$+v^{\gamma}[\langle a^2 \partial^{\alpha}\partial^{\beta}u^{\delta}\rangle+ \langle a^2 \partial^{\alpha}\partial^{\delta}u^{\beta}\rangle+ \langle a^2 \partial^{\beta}\partial^{\delta}u^{\alpha}\rangle]$$
$$+v^{\delta}[\langle a^2 \partial^{\alpha}\partial^{\beta}u^{\gamma}\rangle+\langle a^2 \partial^{\alpha}\partial^{\gamma}u^{\beta}\rangle+ \langle a^2 \partial^{\beta}\partial^{\gamma}u^{\alpha}\rangle]$$

$$+[\langle a^2 u^{\alpha}\partial^{\beta}\partial^{\gamma}u^{\delta}\rangle+ \langle a^2 u^{\alpha}\partial^{\beta}\partial^{\delta}u^{\gamma}\rangle+ \langle a^2 u^{\alpha} \partial^{\gamma}\partial^{\delta}u^{\beta}\rangle]$$
$$+[\langle a^2 u^{\beta}\partial^{\alpha}\partial^{\gamma}u^{\delta}\rangle+ \langle a^2 u^{\beta}\partial^{\alpha}\partial^{\delta}u^{\gamma}\rangle+ \langle a^2 u^{\beta} \partial^{\gamma}\partial^{\delta}u^{\alpha}\rangle]$$
$$+[\langle a^2 u^{\gamma}\partial^{\alpha}\partial^{\beta}u^{\delta}\rangle+ \langle a^2 u^{\gamma}\partial^{\alpha}\partial^{\delta}u^{\beta}\rangle+ \langle a^2 u^{\gamma} \partial^{\beta}\partial^{\delta}u^{\alpha}\rangle]$$
\begin{equation} \label{QPhrt20 T 4 expl p3}
+[\langle a^2 u^{\delta}\partial^{\alpha}\partial^{\beta}u^{\gamma}\rangle+\langle a^2 u^{\delta}\partial^{\alpha}\partial^{\gamma}u^{\beta}\rangle+ \langle a^2 u^{\delta}\partial^{\beta}\partial^{\gamma}u^{\alpha}\rangle]\Biggr].\end{equation}

Approximate equation of state for $T_{s3}^{\alpha\beta\gamma\delta}$
is
$$T_{s3,appr}^{\alpha\beta\gamma\delta}=\frac{\hbar^{2}}{24m^{2}}
\Biggl[nv^{\alpha}[\partial^{\beta}\partial^{\gamma}v^{\delta}+\partial^{\gamma}\partial^{\delta}v^{\beta}+\partial^{\beta}\partial^{\delta}v^{\gamma}]$$
$$+nv^{\beta}[\partial^{\alpha}\partial^{\gamma}v^{\delta}+\partial^{\alpha}\partial^{\delta}v^{\gamma}+\partial^{\gamma}\partial^{\delta}v^{\alpha}]$$
$$+nv^{\gamma}[\partial^{\alpha}\partial^{\beta}v^{\delta}+\partial^{\alpha}\partial^{\delta}v^{\beta}+\partial^{\beta}\partial^{\delta}v^{\alpha}]$$
\begin{equation} \label{QPhrt20 T 4 expl p3 appr} +nv^{\delta}[\partial^{\alpha}\partial^{\beta}v^{\gamma}+\partial^{\alpha}\partial^{\gamma}v^{\beta}+\partial^{\beta}\partial^{\gamma}v^{\alpha}]\Biggr]. \end{equation}
It is equal to zero in the linear approximation for the macroscopically motionless plasmas
since it is nonlinear on the velocity field.

\subsection{Title of the developed approximation}

Various extended hydrodynamics can be developed.
Suggested model is called 20-moment hydrodynamics.
I have five traditional moments: concentration $n$, projections of momentum $nv_{x}$, $nv_{y}$, $nv_{z}$, and energy density (or the temperature) $\varepsilon=p^{\beta\beta}$.
Six functions are in the pressure tensor $p^{\alpha\beta}$, but one of them is taken for the energy density.
Three functions are in the energy current.
Their account leads to the traditional 13-moments approximation.
Furthermore, the symmetric third rank tensor $Q^{\alpha\beta\gamma}$ has 10 independent elements,
but three of them give the energy current.
Therefore, the account of the third rank tensor $Q^{\alpha\beta\gamma}$ makes the model 20-moment hydrodynamics,
where 20-moments are used to describe each species.
The separate spin evolution 20-moment hydrodynamics employs 20 moments for electrons with fixed spin projection.


\begin{thebibliography}{17}


\bibitem{Kremp PRE 99} D. Kremp, Th. Bornath, and M. Bonitz, M. Schlanges, Phys. Rev. E \textbf{60}, 4725 (1999).


\bibitem{Shokri PoP 99} B. Shokri, A. A. Rukhadze,  Phys. Plasmas \textbf{6}, 4467 (1999).

\bibitem{Golubnychiy PRE 01} V. Golubnychiy, M. Bonitz, D. Kremp, and M. Schlanges, Phys. Rev. E \textbf{64}, 016409 (2001).


\bibitem{Haas PRE 00} F. Haas, G. Manfredi, M. Feix, Phys. Rev. E \textbf{62}, 2763 (2000).

\bibitem{Anderson PRE 02} D. Anderson, B. Hall, M. Lisak, and M. Marklund, Phys. Rev. E \textbf{65}, 046417 (2002).


\bibitem{Haas PoP 03} F. Haas, L. G. Garcia, J. Goedert, and G. Manfredi, Phys. Plasmas \textbf{10}, 3858 (2003).


\bibitem{Haas PoP 05} F. Haas, Phys. Plasmas \textbf{12}, 062117 (2005).

\bibitem{Maksimov QHM 99} L. S. Kuz'menkov, S. G. Maksimov,  Theor. Math. Phys. \textbf{118}, 227 (1999).


\bibitem{Shukla UFN 10} P. K. Shukla, B. Eliasson, Phys. Usp. \textbf{53}, 51
(2010) [Uspehi Fizihceskih Nauk \textbf{180}, 55 (2010)].

\bibitem{Shukla RMP 11} P. K. Shukla, B. Eliasson,
Rev. Mod. Phys. \textbf{83},  885 (2011).


\bibitem{MaksimovTMP 2001} L. S. Kuz'menkov, S. G. Maksimov, and V. V. Fedoseev, Theoretical and Mathematical
Physics \textbf{126}, 110 (2001).

\bibitem{MaksimovTMP 2001 b} L. S. Kuz'menkov, S. G. Maksimov, and V. V. Fedoseev, Theor.
Math. Fiz. \textbf{126} 258 (2001) [Theoretical and Mathematical
Physics, \textbf{126} 212 (2001)].


\bibitem{Andreev VestnMSU 2007} P. A. Andreev, L. S. Kuz'menkov,
Moscow University Physics Bulletin \textbf{62}, N.5, 271 (2007).

\bibitem{Marklund PRL07} M. Marklund and G. Brodin,
Phys. Rev. Lett. \textbf{98}, 025001 (2007).

\bibitem{Brodin NJP 07} G. Brodin and M. Marklund, New J. Phys. \textbf{9}, 277
(2007).

\bibitem{Andreev AtPhys 08} P. A. Andreev, L. S.  Kuz'menkov,
Physics of Atomic Nuclei \textbf{71}, N.10, 1724 (2008).

\bibitem{Mahajan PRL 11} S. M. Mahajan and F. A. Asenjo, Phys. Rev. Lett. \textbf{107}, 195003 (2011).
\bibitem{Koide PRC 13} T. Koide, Phys. Rev. C \textbf{87}, 034902 (2013).

\bibitem{Uzdensky RPP 14} D. A. Uzdensky, S. Rightley, Rep. Progr. Phys. \textbf{77}, 036902 (2014).


\bibitem{Andreev PTEP 19} P. A. Andreev, L. S. Kuz'menkov, Prog. Theor. Exp. Phys. \textbf{2019}, 053J01 (2019).

\bibitem{Andreev PoP kinetics 17 a} P. A. Andreev, Phys. Plasmas \textbf{24}, 022114 (2017).

\bibitem{Andreev PoP kinetics 17 b} P. A. Andreev, Phys. Plasmas \textbf{24}, 022115 (2017).


\bibitem{Andreev PoP 16 sep kin} P. A. Andreev, Phys. Plasmas \textbf{23}, 062103 (2016).



\bibitem{Andreev PRE 15 SEAW} P. A. Andreev, Phys. Rev. E \textbf{91}, 033111 (2015).

\bibitem{Andreev AoP 15 SEAW} P. A. Andreev, L. S. Kuz'menkov, Ann. Phys. \textbf{361}, 278 (2015).

\bibitem{Andreev EPL 16} P. A. Andreev, L. S. Kuz'menkov, Eur. Phys. Lett. \textbf{113}, 17001 (2016).
\bibitem{Andreev APL 16} P. A. Andreev, L. S. Kuz'menkov, Appl. Phys. Lett. \textbf{108}, 191605 (2016).


\bibitem{Tokatly PRB 99} I. Tokatly, O. Pankratov,
Phys. Rev. B \textbf{60}, 15550 (1999).


\bibitem{Tokatly PRB 00} I. V. Tokatly, O. Pankratov,
Phys. Rev. B \textbf{62}, 2759 (2000).

\bibitem{Miller PoP 16} S. T. Miller and U. Shumlak, Phys. Plasmas \textbf{23}, 082303 (2016).


\bibitem{Landau Vol X} L. Landau and E. M. Lifshitz, \textit{Statistical Physics, Part II} (Pergamon, New
York, 1980).

\bibitem{Aleksandrov Rukhadze Book} A. F. Aleksandrov, L. S. Bogdankevich, and A. A. Rukhadze,
\textit{Principles of Plasma Electrodynamics}, Berlin; New York: Springer-Verlag, 1984.



%
\end{thebibliography}
\end{document}